\documentclass[12pt]{article}
\usepackage[T1]{fontenc}
\usepackage{lmodern}

\usepackage{amsmath,amsthm,amssymb}
\usepackage[usenames,dvipsnames]{xcolor}
\usepackage{enumerate}
\usepackage{graphicx}
\usepackage{cite}
\usepackage{comment}
\usepackage{oands}
\usepackage{tikz}
\usepackage{changepage}
\usepackage{bbm}
\usepackage{mathtools}
\usepackage[margin=1in]{geometry}
\usepackage{appendix}
\usepackage{stmaryrd}
\usepackage{multicol}
\usepackage{microtype}
\usepackage{dsfont}
\usepackage{mathrsfs}  
\usepackage{multirow}
\usepackage{graphicx,subfigure}
\usepackage{array}
\usepackage{bm}
\usepackage{algpseudocode}
\newcolumntype{P}[1]{>{\centering\arraybackslash}p{#1}}

\usepackage[pdftitle={Luce model},
  pdfauthor={David J. Wei},
colorlinks=true,linkcolor=NavyBlue,urlcolor=RoyalBlue,citecolor=PineGreen,bookmarks=true,bookmarksopen=true,bookmarksopenlevel=2,unicode=true,linktocpage]{hyperref}

\usepackage[capitalise,noabbrev]{cleveref}

\theoremstyle{plain}
\newtheorem{thm}{Theorem}[section]

\def\@rst #1 #2other{#1}
\newcommand\MR[1]{\relax\ifhmode\unskip\spacefactor3000 \space\fi
  \MRhref{\expandafter\@rst #1 other}{#1}}
\newcommand{\MRhref}[2]{\href{http://www.ams.org/mathscinet-getitem?mr=#1}{MR#2}}

\theoremstyle{definition}
\newtheorem{defn}[thm]{Definition}

\numberwithin{equation}{section}

\usepackage{enumitem}
\usepackage{graphicx}
\usepackage{listings}
\usepackage{xcolor}
\usepackage{asymptote}
\usepackage{float}

\usepackage{amsmath,amsthm,amssymb}
\usepackage{algpseudocode}
\usepackage{algorithm}
\usepackage{caption}
\theoremstyle{definition}
\newtheorem{conjecture}{Conjecture}
\title{Uniform meandric systems as percolation on planar maps:\\ conjectures and a numerical study}
\author{David J. Wei}
 \author{
\begin{tabular}{c} David J. Wei\\[-3pt]\small MIT \end{tabular}
}
\date{}
\begin{document}

\maketitle

\begin{abstract}
We perform a numerical study of uniform meandric systems via their encoding by pairs of simple random walks and their interpretation as a model of critical percolation on random planar maps. We investigate certain natural box-crossing events and provide numerical evidence for a strong negative association between opposite open crossings and formulate conjectures on their asymptotic behavior. Our second contribution is a simple law-preserving dynamics on meandric systems, by local peak/valley flips in the walk encoding. Our simulations suggest that the percolation structure decorrelates much faster than the underlying random planar map.
\end{abstract}

\begin{figure}[h!] 
    \centering
    {{\includegraphics[width=0.6\textwidth]{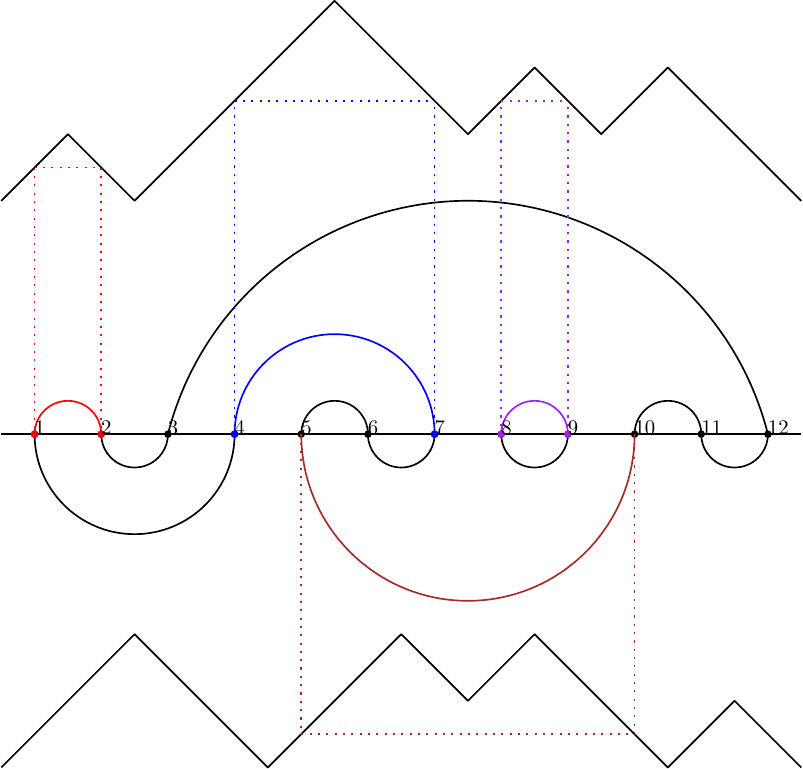}}}%
    \caption{A meandric system of size 6 and the corresponding encoding simple walks are shown (as explained in~\cref{sect:meandric-syst}). Some specific arcs are color-coded to clearly show the one-to-one correspondence between arcs and walk values. Note that the two walks are horizontally translated to the right by $1/2$. Each step of the two walks coincides with an integer point on the real line, as shown above. The reason for this translation is to better illustrate the one-to-one correspondence between the two walks and the meandric system. \label{fig:bijection_with_labels}}%
\end{figure}

\section{Introduction}

A \textbf{meandric system} of size $n$ is a collection of disjoint simple loops in $\mathbb{R}^2$ that cross the horizontal line $\mathbb{R}$ exactly $2n$ times; a \textbf{meander} is a meandric system consisting of a single loop. See \cref{defn:meandric-s} for precise definitions and \cref{fig:bijection_with_labels} for an example. Besides their intrinsic enumerative interest, meanders and meandric systems appear in topology, statistical physics, and free probability, and are closely related to Catalan structures and planar maps; see, for instance,~\cite{ckst-noodle,fn-meander-system,Kargin-cycles-meander,Goulden-Asymptotics-meandric,bgs-meander,Feray-components-mendric,bgp-meander-system,fukuda} and the surveys~\cite{zvonkin-meander-survey,lacroix-meander-survey}.

Following a viewpoint developed in~\cite{bgp-meander-system}, we regard (open) meandric systems as a model of critical percolation on a random planar map, sometimes referred to as a ``quantum lattice''. The basic combinatorial ingredient is the representation of a non-crossing arc diagram by a simple walk excursion, which yields a bijection between (open) meandric systems and pairs of simple walks of the same length; we recall this encoding in \cref{sect:meandric-syst} (see also \cref{fig:bijection_with_labels}). As explained in \cref{sect:perco}, the pair of non-crossing arc diagrams associated with a meandric system canonically induces two non-crossing partitions on interlaced sets of points; see \cref{fig:partitions}. Declaring one family of partition arcs to be ``open'' edges and the other to be ``closed'' edges yields a percolation configuration on a random planar map, whose boundaries are exactly the loops of the meandric system. A key feature of this encoding is that, for each fixed size, the planar map and the percolation configuration determine each other.

In~\cite[Conjecture~1.2]{bgp-meander-system} it is conjectured that, as the size of a uniform meandric system tends to infinity, the associated planar map converges in law to a $\sqrt{2}$-Liouville quantum gravity surface, while the loop ensemble converges to an independent conformal loop ensemble $\mathrm{CLE}_6$. In particular, the planar map and the percolation interfaces should become independent in the scaling limit, even though they are fully coupled in the discrete. The aim of this paper is to provide numerical evidence and a possible mechanism for this phenomenon by studying suitable observables in uniform open meandric systems, both in a static setting and under a natural dynamics which preserves the uniform law.

\medskip

Our first set of results concerns static crossing events in the percolation picture. In \cref{sect:box-crossing}, we introduce box-crossing events for the induced percolation configuration: given a box $B_n(x)=[x,x+2n-1]\times\mathbb{R}$ with $x\in\mathbb{Z}+1/2$, we consider the events that there exists a open path crossing $B_n(x)$ from top-left to bottom-right, and similarly from bottom-left to top-right. For a uniform open meandric system, each of these two events has probability $1/2$ by a simple symmetry argument. In standard Bernoulli percolation on $\mathbb{Z}^2$, the FKG inequality implies that such crossing events are positively correlated. In our setting, numerical experiments indicate the opposite behavior: the existence of a top-left to bottom-right open crossing appears to \emph{hinder} the existence of a bottom-left to top-right open crossing. In \cref{sect:neg-ass}, we therefore formulate two conjectures asserting a strong form of negative association for these two crossing events, see \cref{conj-a} and \cref{conj-b}.

\medskip

Our second contribution is dynamical. In \cref{sect:dynam}, we introduce a simple law-preserving dynamics on uniform open meandric systems. Motivated by the conjectural scaling limit picture and the mating-of-trees description of Liouville quantum gravity~\cite{zbMATH07453876,zbMATH07730640}, we expect that small perturbations of the open meandric systems correspond to small changes of the planar map but may induce macroscopic changes of the percolation interfaces. In \cref{conj1} and \cref{conj2} we predict a separation of time scales for the decorrelation of the percolation structure and the underlying planar map. The numerical implementation of our dynamics and observables is described in detail in \cref{sect:num-sym}. 

\medskip

The paper is organized as follows. In \cref{sect:meandric-syst} we recall the encoding of (open) meandric systems by pairs of walks. In \cref{sect:perco} we review the interpretation of these objects as percolation on a planar map, and in \cref{sect:box-crossing} we introduce the box-crossing events that will be our main observables. \cref{sect:neg-ass} is devoted to the negative association phenomena for crossing events. In \cref{sect:dynam} we define our law-preserving dynamics and state the related conjectures. The numerical setup is presented in \cref{sect:num-sym}, and the outcome of the simulations is discussed in \cref{sect:results}. A brief description of our code is given in the \cref{appendix}.

\bigskip

\noindent\textbf{Acknowledgments.} This work was carried out as part of MIT’s Undergraduate Research Opportunities Program (UROP) under the supervision of Prof.\ Jacopo Borga. The author is deeply grateful to Prof. Jacopo Borga for his insightful guidance and continuous support throughout this research. We also thank Arka Adhikari for some helpful discussions.

\section{Combinatorial constructions}
\subsection{Meandric systems}\label{sect:meandric-syst}

\begin{defn}\label{defn:meandric-s}
A \textbf{meandric system} of size $n$ is a collection $\mathfrak{S}_n$ of finite, simple loops in $\mathbb{R}^2$ satisfying the following three conditions:

\begin{enumerate}
    \item No two loops in $\mathfrak{S}_n$ intersect one another.
    \item Each loop intersects the real line $\mathbb{R}$ at least twice, and no loop intersects $\mathbb{R}$ without crossing it.
    \item In total, the loops in $\mathfrak{S}_n$ cross $\mathbb{R}$ exactly $2n$ times.
\end{enumerate}
\end{defn}
Of course, the $2n$ intersection points between $\mathbb{R}$ and a given system $\mathfrak{S}_n$ can be mapped to the integer points $[1, 2n] \cap \mathbb{Z}$ without changing the structure of the meandric system $\mathfrak{S}_n$ itself. See \cref{fig:bijection_with_labels} for an example.

\medskip

A \textbf{meander} of size $n$ is simply a meandric system consisting of only one loop. 
However, meanders are much more difficult to analyze than meandric systems because meandric systems have a natural bijection with simple walks.

An \textbf{arc diagram} of size $n$ is a collection of arcs in the upper half-plane $\mathbb{R} \times [0, \infty)$ on the set of points $[1, 2n] \cap \mathbb{Z}$ along the real line. There are $n$ arcs in total, and each arc connects exactly two of these points such that no two arcs intersect each other. Then, it is clear that we can view every meandric system as the union of two arc diagrams, one in the upper half-plane and another in the lower half-plane.

Every arc diagram of size $n$ can be bijectively encoded with a \textbf{simple walk excursion} of length $2n$, i.e.\ a simple walk starting at 0, ending at zero and being always non-negative. Such a walk can be represented by a sequence of $2n$ steps $X_1, X_2, \cdots, X_{2n}$, where $X_i \in \{-1, 1\}$ for all $i$. The value of the walk at position $k$, denoted by $S_k$, is given by 
\begin{equation}\label{eq:defm-walk-ms}
    S_k = S_0 + \sum_{i=1}^k X_i,
\end{equation}
where we define $S_0 = 0$ as the starting value of the walk. Then, the bijective encoding is characterized by the following property: two points $x_1, x_2 \in [1, 2n] \cap \mathbb{Z}$ are joined by an arc if and only if \[S_{x_1 - 1} = S_{x_2} < \min_{y \in [x_1, x_2 - 1]} S_y.\] In other words, the value of the walk immediately before the start of an arc at $x_1$ must be equal to the value of the walk at the end of the arc at $x_2$. Moreover, no part of the walk may dip below this value between $x_1$ and $x_2 - 1$. This bijection is illustrated in \cref{fig:bijection_with_labels}.

\begin{figure}[h!] 
    \centering
    {{\includegraphics[width=0.5\textwidth]{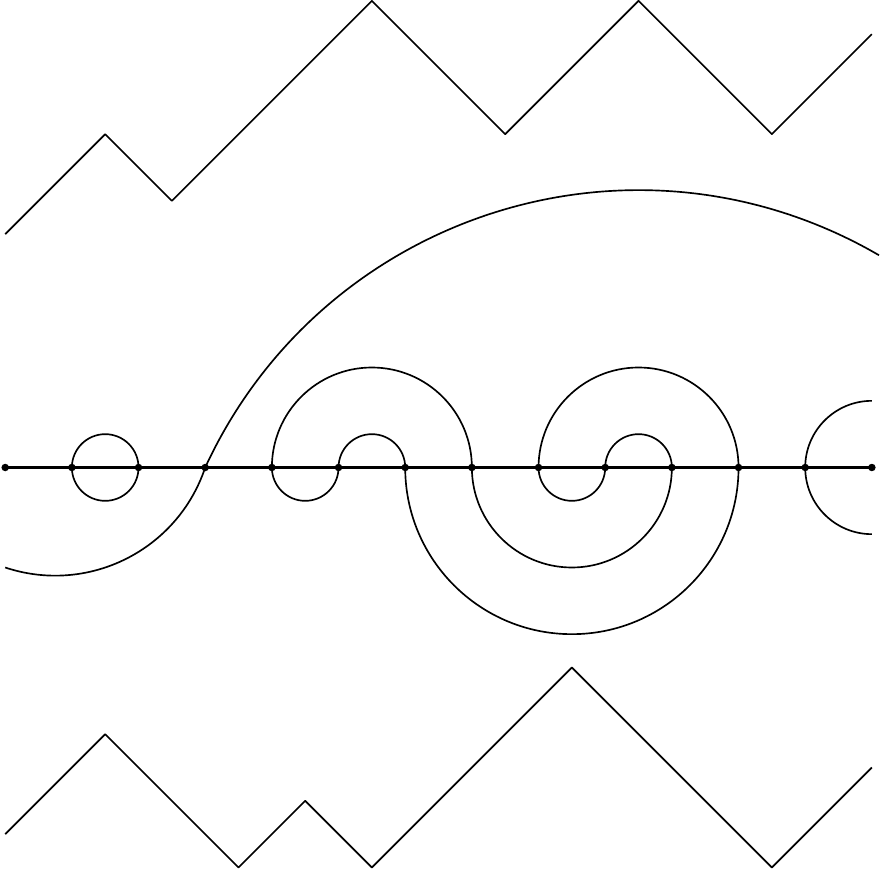}}}%
    \caption{Two simple walks are shown, along with the corresponding open meandric system. \label{fig:open-ms}}%
\end{figure}

Note that if one considers a pair of two \textbf{simple walks} of length $2n$ (not conditioned to be excursion), and runs the same bijection above, one still obtains a collection of non-intersecting loops/curves crossing the line at $2n$ points. We call such a configuration an \textbf{open meandric system} of size $n$. See \cref{fig:open-ms} for an example. 

\subsection{Non-crossing partitions and percolation}\label{sect:perco}

The main goal of this section is to explain how meandric systems can be viewed as a model of percolation on planar maps, as done in \cite[Section 2]{bgp-meander-system}. 

\medskip

We first introduce non-crossing partitions which also play a fundamental role in connections to meandric systems. Formally, a \textbf{non-crossing partition} of a set $S$ is given by a collection $\{C_1, C_2, \cdots, C_k\}$, where $C_i$ is a subset of $S$ for all $1 \le i \le k$, satisfying the following conditions: 

\begin{itemize}
    \item $S = C_1 \cup C_2 \cup \cdots \cup C_k$;
    \item $C_i \cap C_j = \emptyset$ for all $1 \le i \neq j \le k$;
    \item There does \textit{not} exist two distinct subsets $C_i, C_j \subseteq S$ so that we may choose $a_1, a_2 \in C_i$ and $b_1, b_2 \in C_j$ satisfying $a_1 < b_1 < a_2 < b_2$.
\end{itemize}
Intuitively, each point in $S$ belongs to exactly one ``connected component'' of arcs, and no two arcs ever intersect.

A meandric system of size $n$ induces two non-crossing perfect matchings on the points $[1, 2n] \cap \mathbb{Z}$, one in the upper half-plane and the other in the lower half-plane, as shown for instance in  \cite[Section 2.2]{bgp-meander-system}. For now, let us simply focus on the arc diagram in the upper half-plane. Denote this perfect matching by $m$. Then, there exists a unique non-crossing partition $\psi(m)$ on the points $2\mathbb{Z} \cap [1, 2n] - 1/2$ that does not interfere with the arcs in $m$, as described in the next paragraph.

\begin{figure}[H] 
    \centering
    {{\includegraphics[width=0.6\textwidth]{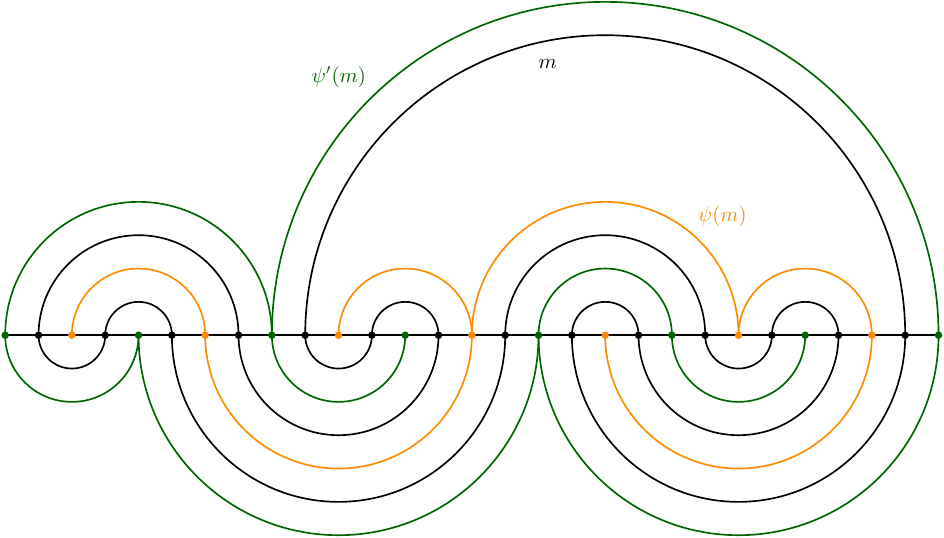}}}%
    \caption{The meandric system $m$ is shown, along with the two non-crossing partitions $\psi(m)$ and $\psi'(m)$ that it induces. The meandric system is formed on the points $\{1, 2, 3, \cdots, 14\}$. Moreover, $\psi(m)$ is formed on the points $\{2 - 1/2, 4 - 1/2, \cdots, 14 - 1/2\}$ while $\psi'(m)$ is formed on the points $\{0 + 1/2, 2 + 1/2, 4 + 1/2, \cdots, 14 + 1/2\}$.\label{fig:partitions}}
\end{figure}

Specifically, there exists a unique non-crossing partition $\psi(m) = \{C_1, C_2, \cdots, C_k\}$ on the shifted lattice $2\mathbb{Z} \cap [1, 2n] - 1/2$ satisfying the following condition: for every pair of distinct partition blocks $C_i$ and $C_j$, and for every pair of points $p \in C_i$ and $q \in C_j$ with $p < q$, there exists a pair of matched points $x_1, x_2 \in m$ satisfying either $x_1 < p < x_2 < q$ or $p < x_1 < q < x_2$. See \cref{fig:partitions} for an example.

Visually, the idea is that we draw arcs between points in each partition block $C_i$ without having any arcs crossing the arcs already present in the perfect matching $m$. 

Similarly, we notice that there exists a unique non-crossing partition $\psi'(m)$ on the points $2\mathbb{Z} \cap [0, 2n] + 1/2$, following the same rules we just described.

Thus, every meandric system, which can be represented as a pair of non-crossing matchings $m = (m_1, m_2)$, effectively induces two pairs of non-crossing partitions, $\psi(m_1) \cup \psi(m_2)$ and $\psi'(m_1) \cup \psi'(m_2)$. An example is shown in \cref{fig:partitions}, with the green arcs showing one non-crossing partition and the orange ones showing another. From now on, we will refer to the green points (resp.\ orange points) as the points $2\mathbb{Z} \cap [0, 2n] + 1/2$ (resp.\ $2\mathbb{Z} \cap [1, 2n] - 1/2$). Similarly, we will let the green arcs (resp.\ orange arcs) refer to the arcs associated with the non-crossing partitions on the green points (resp.\ orange points). 

\paragraph{Meandric systems as a model of percolation on a planar map.} The meandric system $m$ and its two non-crossing partitions form a percolation model. We may draw analogies with classical quantities for (more standard) percolation models. For example, we can view the union of green and orange points (the points $\mathbb{Z} \cap [0, 2n] + 1/2$) as the vertices of our planar map, and the union of the green and orange arcs, along with the segments connecting adjacent green and orange points, as the edges of our planar map. Moreover, without loss of generality, we may think of the green edges as \emph{open edges} and the orange edges as \emph{closed edges}. The black loops formed by the black arcs in the perfect matchings in the meandric system can be thought of as \emph{boundaries of clusters} of open edges (see again \cref{fig:partitions}). We stress the peculiar fact that if one knows the planar map it can uniquely determine the boundaries of open clusters, and vice-versa. That is, the planar map and the boundaries of open clusters determine each other.

\subsection{Box crossings}\label{sect:box-crossing}
Here, we will work in the space $\mathbb{R}^2$ and assume that all the arcs are drawn as semi-circles.  We define a box $B_n(x)$ as a box with width $2n - 1$, starting at $x$ and ending at $x + 2n - 1$.  In other words, we let $B_n(x) := [x, x + 2n - 1] \times \mathbb{R}$. Consider a meandric system $m = (m_1, m_2)$, where $m_1$ and $m_2$ are the non-crossing matchings in the upper and lower half-planes, respectively.  This system induces two pairs of non-crossing partitions as described earlier (the collection of green and orange arcs).

\begin{figure}[H] 
    \centering
    {{\includegraphics{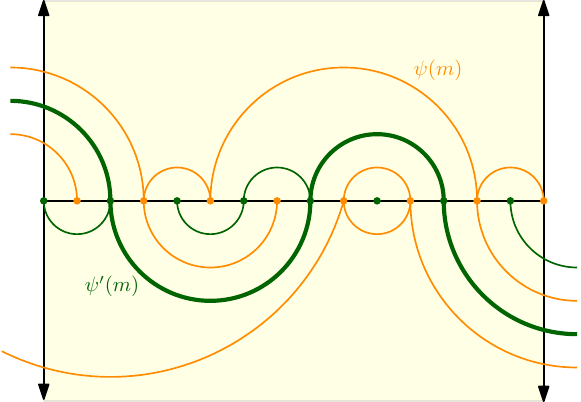}}}%
    \caption{The green points are at positions $\{0+1/2, 2+1/2, 4+1/2, \cdots, 14+1/2\}$, while the orange points are at positions $\{2-1/2, 4-1/2, \cdots, 16-1/2\}$. Note that the meandric system $m$ itself is not shown. Only the two induced non-crossing partitions $\psi(m)$ and $\psi'(m)$ are shown. Given our definitions, we are working with box $B_8(1/2)$. The top-left to bottom-right green crossing is bolded. \label{fig:box}}%
\end{figure}

Now, we define crossing events similar to those in standard Bernoulli percolation. We say that an arc crosses the top-left boundary $\{x\} \times [0, \infty)$ of a box $B_n(x)$ if the edge touches this vertical line (note that it includes the point $(x,0)$). Similarly, we say that an edge crosses the bottom-right boundary $\{x+2n-1\} \times (-\infty, 0]$ if the edge touches this vertical line (note that it includes the point $(x+2n-1,0)$). Then, we say that there exists a top-left to bottom-right green path (resp.\ orange path) inside this box if there exists a sequence of connected all green (resp.\ orange) edges, where the first edge in the sequence crosses the top-left boundary of $B_n(x)$ and the last edge in the sequence crosses the bottom-right boundary of $B_n(x)$. Other crossing events, like bottom-left to top-right crossings, are defined similarly. An example of a top-left to bottom-right green crossing is shown in \cref{fig:box}. 
Given a (large enough) uniform open meandric system and a box $B_n(x)$ with $x\in \mathbb{Z}+1/2$, let $A$  denote the event of a top-left to bottom-right green crossing, and let $B$ denote the event of a orange bottom-left to top-right crossing. A standard observation in percolation theory is that exactly one of the events $A$ and $B$ can hold. Hence, using a symmetry argument, one can precisely determine that 
\begin{equation}\label{eq:1/2-proba}
    \mathbb{P}(A) = \mathbb{P}(B) = 1/2,
\end{equation}
as shown in \cite[Proposition 2.1]{bgp-meander-system}.

\subsection{Peaks and valleys}
It will be useful to define the notion of peaks and valleys in simple walks. Given a sequence of $2n$ steps $X_1, X_2, \cdots, X_{2n} \in \{-1, 1\}$, let $S_k$ denote the value of the walk at position $k$, as defined in \eqref{eq:defm-walk-ms}, where $0 \le k \le 2n$.

Then, we say that the walk has a \textit{peak} at position $k$ if one of the following three conditions is satisfied:

\begin{itemize}
    \item $k = 0$ and $S_0 > S_1$;
    \item $k = 2n$ and $S_{2n} > S_{2n - 1}$;
    \item $1 \le k \le 2n-1$ and $S_k > S_{k-1}$ and $S_k > S_{k+1}$.
\end{itemize}
Similarly, $S_k$ is a \textit{valley} if one of the following three conditions is satisfied:
\begin{itemize}
    \item $k = 0$ and $S_0 < S_1$;
    \item $k = 2n$ and $S_{2n} < S_{2n - 1}$;
    \item $1 \le k \le 2n-1$ and $S_k < S_{k-1}$ and $S_k < S_{k+1}$.
\end{itemize}

\section{Negative association in meandric systems}\label{sect:neg-ass}

Recall that two events $A$ and $B$ are positively correlated if $P(A \cap B) \ge P(A)P(B)$. If the inequality sign is flipped, we say that the two events are negatively correlated.

For standard Bernoulli bond percolation on $\mathbb{Z}^2$ (and many other variants), it is well known that two \emph{increasing} events are always positively correlated, thanks to the FKG inequality~\cite[Chapter 2.2]{zbMATH01301486}. As a canonical example of increasing events, we may consider the event $A$ of a left to right crossing of an $n\times n$ box, which is a sequence of connected open edges from the left boundary of the $n \times n$ grid to the right boundary, and the event $B$ of a top to bottom crossing of the same $n\times n$ box, which is defined similarly. Then, thanks to the FKG inequality, these two events are positively correlated. Intuitively, the existence of a left to right crossing allows for more open edges, which would potentially help the existence of a top to bottom crossings.

We previously saw how meandric systems can be connected to percolation theory, so one might expect that a similar positive association exists for meandric systems. However, we believe this is not the case, at least in our interpretation of meandric systems as a model of percolation in planar maps. Indeed, if we consider the crossing events introduced in \cref{sect:box-crossing},  then the existence of a top-left to bottom-right green crossing \emph{does not} increase the chances of the existence of a bottom-left to top-right green crossing. 

\begin{conjecture}\label{conj-a}
For every $n\geq 1$ and $x\in \mathbb{Z}+1/2$, given a box $B_n(x)$ and a uniform open meandric system, the probability that there exists both a top-left to bottom-right green crossing and a bottom-left to top-right green crossing is bounded above by $1/4$.
\end{conjecture}

The bound of $1/4$ is derived from the fact that the probability of each green crossing separately is precisely $1/2$, recall \eqref{eq:1/2-proba}; and if the two green crossings are indeed negatively correlated, then the probability of both green crossings should be no larger than $1/2 \times 1/2 = 1/4$.

In fact, from numerical simulations (see Tables~\ref{table1}~and~\ref{table2}), the existence of one of the two green crossings hinders the existence of the other green crossing, suggesting a very strong negative correlation between these two crossing events in the meandric system.

\begin{conjecture}\label{conj-b}
Consider a box $B_n(x)$ with $x\in \mathbb{Z}+1/2$ and a uniform open meandric system. The probability that there exists both a top-left to bottom-right green crossing and a bottom-left to top-right green crossing monotonically decreases to zero as $n$ tends to infinity.
\end{conjecture}

In other words, as the size of the box we are considering increases, numerical simulations suggest that the probability that there exists both green crossings decreases.

For small values of $n$, we were able to generate all possible systems by considering all possible $2^{2n-1} \times 2^{2n-1} = 2^{4n-2}$ pairs of simple walks. We computed the number of systems that have both green crossings, as well as the number of systems in total, as shown in Table~\ref{table1}.
\begin{table}[h!]
\begin{center}
\begin{tabular}{c|c|c|c|c|c|c}\label{exact_probs}
Box size & $1$ & $2$ & $3$ & $4$ & $5$ & $6$ \\
\hline
Both crossings & $1$ & $14$ & $208$ & $3,158$ & $48,520$ & $751,112$ \\
\hline
Total & $4$ & $64$ & $1,024$ & $16,384$ & $262,144$ & $4,194,304$ \\
\hline
Probability & $0.25$ & $0.22$ & $0.20$ & $0.19$ & $0.185$ & $0.179$
\end{tabular}
\end{center}
\caption{The table shows the number of meandric systems that have both green crossings, as well as the approximate probabilities that a randomly selected meandric system of that size wil have both crossings.}
\label{table1}
\end{table}

For larger values of $n$, we numerically estimated the probabilities, as shown in Table~\ref{table2}.
\begin{table}[h]
\begin{center}
\begin{tabular}{c|c|c|c|c|c}
Box size & $10$ & $50$ & $100$ & $500$ & $1000$ \\
\hline
Probability both crossings & $0.16$ & $0.13$ & $0.11$ & $0.093$ & $0.083$
\end{tabular}
\end{center}
\caption{The table shows the probabilities of having both crossings for larger sized boxes when a meandric system of that size is randomly selected.}
\label{table2}
\end{table}

\section{A law-preserving dynamics for uniform meandric systems}\label{sect:dynam}

In \cref{sect:perco}, we saw how to interpret meandric systems as a model of percolation on a planar map. In \cite[Conjecture 1.2]{bgp-meander-system} it is conjectured that if one considers a uniform meandric system of size $n$, the scaling limit as $n$ tends to infinity of the planar map and the boundaries of the open clusters (black loops) should become independent (in particular, the planar map should converge to $\sqrt{2}$-Liouville quantum gravity surface and the percolation model should converge to an independent copy of a conformal loop ensemble of parameter $6$). See \cite{bgp-meander-system} for more details on these objects.

One of the main difficulties in approaching this conjecture is that for every fixed $n$, as already mentioned at the end of Section~\ref{sect:perco}, the planar map and the  boundaries of the open clusters determine each other (but are believed to become independent in the limit). Our goal in this section is to try to give a possible explanation for this phenomenon. Our idea is to show that if one slightly perturbs the planar map (so that the scaling limit remains unchanged) then it produces a macroscopic change on the boundaries of the open clusters (so that the scaling limit becomes independent from the unperturbed one). We will perturb meandric systems by introducing a simple dynamics on uniform meandric systems that preserves their law. 
Our next goal is to introduce such dynamics.

\paragraph{A law-preserving dynamics on uniform meandric systems.} Consider a uniform open meandric system $m$ of size $n$ represented by the pair of simple random walks $(S_\text{top}, S_\text{bot})$ of length $2n$. 
We select a position $p \in \{1, 2, \cdots, 2n\}$ uniformly at random, and then we select either the top or bottom walk (again uniformly at random). If we choose the top walk, and $p$ is either a peak or a valley in $S_\text{top}$, then we flip it (otherwise we do nothing). In other words, if $p$ is initially a peak, it becomes a valley in $S_\text{top}$, and if $p$ is initially a valley, then it becomes a peak in $S_\text{top}$. The same thing happens if we select the bottom walk. Thus, each step of this dynamics will potentially affect exactly one position in exactly one of the two walks. We denote by $(S^t_\text{top}, S^t_\text{bot})$ the walks obtained after $t$ iterations of this dynamics and by $m^t$ the corresponding open meandric system.

Clearly, this dynamics preserves the law of the uniform open meandric system since it preserves the law of the two walks. We want to investigate how the planar map and the boundaries of the open clusters change under this dynamics. To do this, we focus on the mixing properties of the following two quantities:

\begin{itemize}
    \item The final values $S_\text{top}(2n)$ and $S^t_\text{top}(2n)$ of the top random walk before and after $t$-iterations of the dynamics.

    \item The events $A$ and $A^t$ that there exist a top-left to bottom-right green box crossings in $m$ and $m_t$ (i.e.\ before and after $t$-iterations of the dynamics), respectively.
\end{itemize}

Our goal will be to numerically verify (see Section~\ref{sect:results} for the results) the following conjecture:

\begin{conjecture}\label{conj1}
    There exists some $t=t(n)$ such that 
    \[c_n(t):=\frac{1}{2n}\mathbb{E}[S_\text{top}(2n) \cdot S^t_\text{top}(2n)]\xrightarrow[n\to \infty]{}1\]
    but
    \[p_n(t):=\mathbb{P}(A\cap A^t)\xrightarrow[n\to \infty]{}\mathbb{P}(A)\cdot\mathbb{P}(A^t)=1/4,\]
    that is, after $t=t(n)$ steps of the dynamics, the final value of the top random walk remains highly correlated with its original final value, while the box crossing events before and after the dynamics become asymptotically independent.
\end{conjecture}

More precisely we believe that the following fact holds:

\begin{conjecture}\label{conj2}
    There exists $\alpha,\beta>0$ with $\alpha<\beta$ such that for all $0<\varepsilon<\min\{\alpha,\beta\}$:
    \begin{itemize}
    \item If $t=t(n)=n^{\beta-\varepsilon}$, 
    \[c_n(t):=\frac{1}{2n}\mathbb{E}[S_\text{top}(2n) \cdot S^t_\text{top}(2n)]\xrightarrow[n\to \infty]{}1,\]
    but if $t=t(n)=n^{\beta+\varepsilon}$, 
    \[c_n(t):=\frac{1}{2n}\mathbb{E}[S_\text{top}(2n) \cdot S^t_\text{top}(2n)]\xrightarrow[n\to \infty]{}0.\]
    \item If $t=t(n)=n^{\alpha-\varepsilon}$,
    \[p_n(t)=\mathbb{P}(A\cap A^t)\xrightarrow[n\to \infty]{}\mathbb{P}(A)=1/2,\]
    but if $t=t(n)=n^{\alpha+\varepsilon}$,
    \[p_n(t)=\mathbb{P}(A\cap A^t)\xrightarrow[n\to \infty]{}\mathbb{P}(A)\cdot\mathbb{P}(A^t)=1/4,\]
    \end{itemize}
    that is, the decorrelation time for the ending point of the walk is of order $n^{\beta}$ but the decorrelation time for the crossing events is of order $n^{\alpha}$, that is, crossing events decorrelate much faster.
\end{conjecture}

\medskip

We finally give a heuristic explanation of why these two statistics should be strictly related to the changes (under the dynamics) of the planar map and of the boundaries of the open clusters.
As mentioned earlier, the scaling limit of the planar map should be given by $\sqrt{2}$-Liouville quantum gravity. In particular, thanks to the mating-of-trees encoding of Liouville quantum gravity~\cite{zbMATH07453876,zbMATH07730640}, we believe that the Brownian motion $(X,Y)$, describing the scaling limit of $(S_\text{top}, S_\text{bot})$, should encode such Liouville quantum gravity, but should \emph{not} encode any information of the limiting conformal loop ensemble. See \cite[Section 1.2]{bgp-meander-system} for more explanations. This is why we believe that a small change in the final value of the top random walk should be related to a small change in the scaling limit of the planar map. On the other hand, it is natural to believe  that a big change in the existence of a top-left to bottom-right green box crossings should be related to a big change in the scaling limit of the boundaries of the open clusters.

\section{Implementation of the numerical simulations}\label{sect:num-sym}

We now describe how we implemented our numerical simulations.

We fix a positive integer $n$ denoting the size of the box $B_n(1/2)$ we  consider. We then sample $N=1000$ open meandric systems 
\[m^0_i,\qquad\text{for $i\in\{1,\dots,N\}$},\]
in the box $B_n(1/2)$ by sampling two simple random walks $(S^0_{\text{top},i}, S^0_{\text{bot},i})$ and using the bijection in \cref{sect:meandric-syst}.
Finally,  we apply $t$-times the dynamics of  \cref{sect:dynam}, obtaining the meandric systems $m^t_i$ corresponding to the walks $(S^t_{\text{top},i}, S^t_{\text{bot},i})$.
For each $t\geq 0$, we then compute the quantities:

\[c_n(t) \approx \dfrac{1}{2n} \left[\dfrac{1}{N}\sum_{i=1}^N \Big(S^0_{\text{top},i} (2n)\cdot S^t_{\text{top},i} (2n)\Big)\right]
\quad\text{and}\quad p_n(t) \approx \dfrac{1}{N} \sum_{i=1}^N \mathds{1}_{A^0_i\cap A^t_i},\] 
where $A^t_i$ is the event that there is a top-left to bottom-right crossing in the open meandric system $m_i^t$.

\section{Numerical results}\label{sect:results}
\begin{figure}[H] 
    \centering
    {{\includegraphics[width=0.8\textwidth]{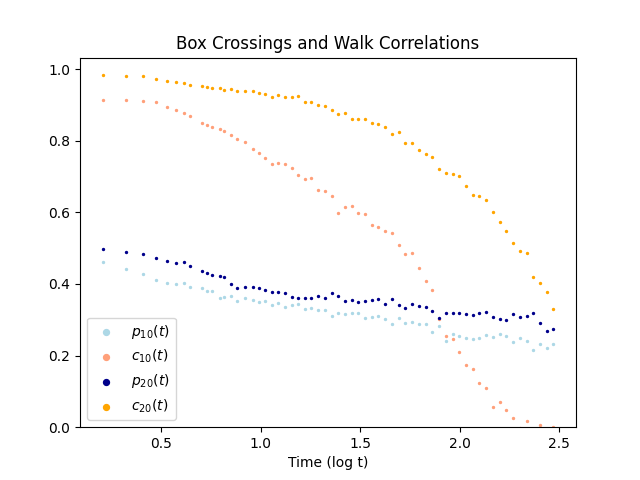}}}%
    \caption{The crossing probabilities and walk correlations of meandric systems of sizes $n=10$ and $n=20$.\label{fig:mixing}}
\end{figure}

The graph in \cref{fig:mixing} illustrates the quantities $p_n(t)$ and $c_n(t)$, when the box $B_n(1/2)$ has sizes $n = 10$ and $n = 20$. We have plotted the graphs with $\log t$ on the $x$-axis.

\noindent The blue points represent the estimated probabilities $p_{n}(t)$ that there exists a top-left to bottom-right crossing both at time 0 and time $t$. From both the $n=10$ and $n=20$ figures, we see that the probabilities initially start at $1/2$, as expected. As $t$ increases to a sufficiently large value, the probability converges to $1/4$.

The orange points represent the correlation $c_n(t)$. Initially, the correlations are close to one, as expected. As $t$ becomes sufficiently large, the correlation value drops to zero, again as expected. 
We also observe that the blue dots initially stay relatively constant at $0.5$ before the probability begins to converge towards $0.25$. 

\medskip

More specifically, we estimated that the time it would take for the dynamics so that the box crossing probability drops below a certain threshold to be of order $n^\alpha$ for some exponents $\alpha$. We observed that the amount of time $t$ it would take  
for $p_n(t)$ to drop below $0.375$ is of order $ n^{\alpha_1}$, and the amount of time it would take for $p_n(t)$ to drop below $ 0.25 + \varepsilon$ to be of order $ n^{\alpha_2}$, where we set $\varepsilon = 0.05$. Tables~\ref{table3}~and~\ref{table4} show the numerical values of $\alpha_i$ for $i = 1, 2$ for differently sized boxes. Since we are assuming $t \sim n^{\alpha_i}$, we have the relation $\log t \approx \alpha_i \log n + c$ for some constant $c$. Thus, to compute $\alpha_i$, we perform a linear regression of $\log t$ onto $\log n$ and extracted the coefficient of $\log n$.


\begin{table}[h]
\begin{center}
\begin{tabular}{c|c|c|c|c|c|c}
Box size $n$ & $2^2$ & $2^3$ & $2^4$ & $2^5$ & $2^6$ & $2^7$ \\
\hline
minimal $t$ with $p_n(t) < 0.375$ & $5$ & $13$ & $47$ & $150$ & $384$ & $1338$
\end{tabular}
\qquad $\alpha_1 \approx 1.62$
\caption{Time $t$ for which $p_n(t) < 0.375$ with exponentially increasing box size.}
\label{table3}
\end{center}
\end{table}

\begin{table}[h]
\begin{center}
\begin{tabular}{c|c|c|c|c}
Box size $n$ & $2^2$ & $2^3$ & $2^4$ & $2^5$ \\
\hline
minimal $t$ with $p_n(t) < 0.25 + \varepsilon$ & $22$ & $139$ & $611$ & $3115$
\end{tabular}
\qquad $\alpha_2 \approx 2.36$
\caption{Time $t$ for which $p_n(t) < 0.25+\varepsilon$ with exponentially increasing box size.}
\label{table4}
\end{center}
\end{table}

Similarly, we also estimated that the time it would take for the correlation $c_n(t)$ to drop below a certain threshold value to be of order $n^\beta$ for some exponent $\beta>0$. We observed that the amount of time $t$ it would take for $c_n(t)$ to drop below $ 0.5$ to be of order $n^{\beta_1}$, and the amount of time it would take for $c_n(t) $ to drop below $ \varepsilon$ to be of order $ n^{\beta_2}$, where  $\varepsilon = 0.05$. Tables~\ref{table5}~and~\ref{table6} show the simulated values of $\beta_i$ for $i = 1, 2$ for differently sized boxes.


\begin{table}[h!]
\begin{center}
\begin{tabular}{c|c|c|c|c}
Box size $n$ & $2^2$ & $2^3$ & $2^4$ & $2^5$ \\
\hline
$t$ with $c_n(t) < 0.5$ & $7$ & $117$ & $1054$ & $10162$
\end{tabular}
\qquad $\beta_1 \approx 3.47$
\end{center}
\caption{Time $t$ for which $c_n(t) < 0.5$ with exponentially increasing box size.}
\label{table5}
\end{table}

\begin{table}[h!]
\begin{center}
\begin{tabular}{c|c|c|c}
Box size $n$ & $2^2$ & $2^3$ & $2^4$ \\
\hline
$t$ with $c_n(t) < \varepsilon$ & $76$ & $776$ & $6603$
\end{tabular}
\qquad $\beta_2 \approx 3.22$
\end{center}
\caption{Time $t$ for which $c_n(t) < \varepsilon=0.05$ with exponentially increasing box size.}
\label{table6}
\end{table}

Note that we have less data for the correlations $c_n(t)$ than for the probabilities $p_n(t)$ because our simulations in Figure~\ref{fig:mixing} show that the amount of time the dynamics take to reach a small value of $c_n(t)$ is very large.

The fact that both $\alpha_i$ are much smaller than both $\beta_i$ provides strong support for \cref{conj1} and \cref{conj2}.

\appendix

\section{A few comments on our code}\label{appendix}

Suppose our program described in \cref{sect:num-sym} runs for $T$ iterations of the dynamics. Then, the overall structure of our program is as in \cref{alg:cap}. Our code is available at: \url{https://github.com/davidw289/Meandric-System-Numerical-Analysis}. 

\begin{algorithm}[h!]
\caption{Simulation}\label{alg:cap}
\begin{algorithmic}
    \State $\texttt{paths} \gets \emptyset$
    \State $\texttt{initial\_crossing} \gets \emptyset$
    \State $\texttt{initial\_endpoint} \gets \emptyset$

    \For{$k \gets 1$ to $N$} \Comment{Generate $N$ random meandric systems/boxes}
        \State $p \gets \texttt{GenerateSystem}(n)$
        \State Unpack $(\texttt{top}, \texttt{bottom}) \gets p$
        \State append $p$ to \texttt{paths}
        \State append $\texttt{CheckPath}(\texttt{top},\texttt{bottom})$ to \texttt{initial\_crossing}
        \State append $\texttt{top}(2n)$ to \texttt{initial\_endpoint}
    \EndFor

    \State $(P_\text{list}, C_\text{list}) \gets (\emptyset, \emptyset)$
    \For{$t \gets 0$ to $T$}
        \State $\texttt{counter} \gets 0$
        \State $\texttt{correlation\_sum} \gets 0$

        \For{$i \gets 1$ to $N$}
            \State Unpack $(\texttt{top}, \texttt{bottom}) \gets \texttt{paths}[i]$

            \If{$t > 0$}
                \State Choose $r \sim \operatorname{Uniform}(0,1)$
                \Comment{Randomly choose top or bottom path}
                \State Choose $j \sim \operatorname{Uniform}(\{1,2,\ldots,2n\})$
                \Comment{Randomly choose an index in the system}

                \If{$r > 0.5$ and $j \in \texttt{GetExtrema}(\texttt{top})$}
                    \State flip top path at index $j$
                \ElsIf{$r \leq 0.5$ and $j \in \texttt{GetExtrema}(\texttt{bottom})$}
                    \State flip bottom path at index $j$
                \EndIf

                \State $\texttt{paths}[i] \gets (\texttt{top},\texttt{bottom})$
            \EndIf

            \If{$\texttt{initial\_crossing}[i]$ and
            $\texttt{CheckPath}(\texttt{top},\texttt{bottom})$}
                \Comment{Check for a crossing at both times $0$ and $t$}
                \State $\texttt{counter} \gets \texttt{counter}+1$
            \EndIf

            \State $\texttt{correlation\_sum}
            \gets \texttt{correlation\_sum}
            +\texttt{initial\_endpoint}[i]\texttt{top}(2n)$
        \EndFor

        \State $q \gets \texttt{counter}/N$
        \State $c \gets \texttt{correlation\_sum}/(2nN)$
        \State append $q$ to $P_\text{list}$
        \State append $c$ to $C_\text{list}$
    \EndFor
\end{algorithmic}
\end{algorithm}

The $\texttt{CheckPath}$ function computes whether there is a top-left to bottom-right crossing in the meandric system. This function is the most delicate part of our code since we had to optimize it quite carefully in order to be able to handle large simulations.

\newpage

\bibliography{cibib,cibib2,references}
\bibliographystyle{hmralphaabbrv}

\end{document}